\documentclass[preprint,proceedings]{rmaa}

\usepackage{paralist}

\usepackage{psfrag,color}

\SetYear{2003}
\SetConfTitle{Environments \&\ Evolution of Double \&\ Multiple Stars, IAU Coll.191}

\title{Spectroscopy, Photometry and Micro-arcsec Astrometry of Binaries
with the {\sl GAIA} Space Mission and with the {\sl RAVE} Experiment} 

\author{
  Toma\v{z} Zwitter,\altaffilmark{1} 
  and Ulisse Munari\altaffilmark{2}
}

\altaffiltext{1}{University of Ljubljana, Department of Physics, Ljubljana, Slovenia.}
\altaffiltext{2}{Osservatorio Astronomico di Padova, Sede di Asiago, Asiago (VI), Italy.}

\shortauthor{Zwitter \& Munari}
\shorttitle{Binaries with {\sl GAIA} and {\sl RAVE}}

\fulladdresses{
\item Ulisse Munari: Osservatorio Astronomico di Padova, Sede di Asiago, 
I-36012 Asiago (VI), Italy (\email{munari@pd.astro.it}).
\item Toma\v{z} Zwitter: University of Ljubljana, Department of Physics, Jadranska 19,
1000 Ljubljana, Slovenia (\email{tomaz.zwitter@fmf.uni-lj.si}).
}

\listofauthors{T. Zwitter, \& U. Munari}
\indexauthor{Zwitter, T.}
\indexauthor{Munari, U.}

\abstract{GAIA astrometric mission of ESA will be very efficient in discovering 
binary and multiple stars with any orbital period, from minutes to 
millions of years. Main parameters of the revised mission design are presented. Next 
we estimate the fraction of binary stars discovered by means of astrometry, 
photometry and on-board spectroscopy. Finally we summarize observations that 
confirm the ability to measure physical parameters like masses, radii and spectroscopic
distance from GAIA data alone. GAIA will fly only in 2010, but the Radial velocity 
experiment (RAVE) has started this year. We show that its spectroscopic observations have 
the capacity to discover a large fraction of so far unknown binary systems. 
}

\addkeyword{binaries: eclipsing}
\addkeyword{binaries: spectroscopic}
\addkeyword{stars: fundamental parameters}
\addkeyword{surveys: GAIA}
\addkeyword{surveys: RAVE}

\begin{document}
\maketitle

\section{Introduction}

{\sl Hipparcos} astrometric mission of ESA in the early nineties had a 
remarkable success. For 118,218 targets listed in the Hipparcos 
catalogue it was able to obtain good precision astrometry 
(10\%\ error at 100~pc), accurate proper motions (error of 1~mas$/$yr) 
and a three colour photometry. 
The satellite scanned the sky, each object was observed in $\sim 30$ 
independent geometrical observations, so a simple five-parameter model 
could be expanded to discover a number of new double and multiple stars. 
Actual number of observations of each target was even larger ($\sim 110$)
so a number of variable stars as well as eclipsing binaries was discovered 
from Hipparcos light curves (ESA SP-402, vol.~12). 

\begin{table*} \centering
\setlength{\tabnotewidth}{0.9\textwidth}
\tablecols{4}
\caption{Summary of capabilities of Hipparcos, SIM and GAIA missions
\tabnotemark{a} }
\begin{tabular}{llll}
\toprule
         & Hipparcos & SIM & GAIA \\ \midrule
agency   & ESA       & NASA& ESA \\
mission lifetime& 4 yrs& 5 yrs & 5 yrs \\
launch   & 1989      & end of 2009& 2010 \\
No. of stars& 120,000& 10,000& 1 billion \\
mag. limit & 12      & 20   & 20 \\
astrometric accuracy &
             1~mas (at V$=10$)& 3$\mu$as (at V$=20$) & 3$\mu$as (at V$=12$)\\
	 &            &     & 10$\mu$as (at V$=15$)\\
	 &            &     &200$\mu$as (at V$=20$)\\
photometric filters& 3 (BBP) & & 5 (BBP\tabnotemark{b}), up to 16 (MBP\tabnotemark{c})\\
radial velocity &
                  not available & not available & $\sigma \sim$1 km~s$^{-1}$ (at $\mathrm{I_C}=14$)\\
                &     &     &$\sigma \sim $10 km~s$^{-1}$ (at $\mathrm{I_C}=16$)\\
epochs on each target& $\sim 110$&    pointed  & $\sim 82$ (astrometry, BBP\tabnotemark{b})\\
                      &          &             & $\sim 200$ (MBP\tabnotemark{c})\\
		      &          &             & $\sim 100$ (spectroscopy) \\
\bottomrule
\tabnotetext{a}{Adapted from ESA-SCI(2000)4, Jordi et al.\ (2003), and Munari et al.\ (2003).}
\tabnotetext{b}{broad-band photometry.}
\tabnotetext{c}{medium-band photometry.}
\end{tabular}

\end{table*}

All these results prompted preparations for its successor even before the 
Hipparcos observations were completed. Already in 1995 ESA  embraced 
the idea of the GAIA mission (ESA SP-379). It was later approved as a Cornerstone 6 mission
and in May 2002 it was re-approved within the new Cosmic Vision 2020 
programme of ESA to fly around 2010. GAIA will improve the results of 
Hipparcos in many areas (see Table~1). Its astrometry will be of a much 
higher precision and reaching for fainter targets. Astrometry will be used 
to obtain 5 coordinates in space-velocity of each object, with an 
on-board spectroscopy supplying the missing radial velocity. Accurate 
multi-band photometry will be used to assess chromaticity corrections 
and to judge basic parameters of stellar atmospheres like 
temperature, gravity and metallicity. The plan is to observe any of the 
$\sim$ billion stars brighter than $V=20$ repeatedly when the scanning 
satellite will bring it into the field of view. This is very different 
from high-precision but pointed observations of a moderate number of 
targets to be observed by the SIM mission. Some smaller photometric missions 
(COROT, Eddington) are discussed elsewhere in this volume (Maceroni 2003). 

Scientific drivers and technical solutions of the GAIA mission have been 
extensively described in the ESA's Concept and Technology Study (ESA-SCI(2000)4)
and in papers of Gilmore et al.\ (1998) and Perryman et al.\ (2001). Several 
conferences have been devoted to definition of its scientific goals 
(Strai\v{z}ys 1999, Bienaym\`{e} \&\ Turon 2002, Vansevi\v{c}ius et al.\@ 2002, 
and Munari 2003). Here we will start with a non-technical description of 
the scientific payload of the satellite and continue with assessment of 
GAIA's capabilities for discovery and modeling of binary and multiple stellar 
systems. In the end we will briefly describe the ground-based project RAVE 
which is about to begin collecting radial velocities and near-IR spectra for 
the brighter end of the GAIA objects within this year. 
\medskip

\section{GAIA astrometry}

GAIA astrometry builds on the concept proven by the Hipparcos satellite. Two astrometric 
telescopes are used. They are pointing in directions perpendicular to the telescope axis 
which are $106^\mathrm{o}$ apart. The image of both telescopes forms in a common focal 
plane. The satellite rotates around its axis every 6~hours, so the stars drift along the 
focal plane. The plane is covered with an array of 110 astrometric CCDs read in a 
time-delayed-integration mode, 
so that angles along the drift direction between stars seen by the first and the second 
telescope can be measured with extreme accuracy. The satellite axis precesses in 70 days along the 
cone with a half opening angle of $50^\mathrm{o}$ around the direction pointing away from 
the Sun. Combination of  rotation, precession and motion around the Sun guarantees that each 
point in the sky is observed in many epochs during the 5 year mission lifetime. 

Extreme astrometric precision requires very stable observing conditions. The whole structure 
will be made of SiC, but even so the temperature variations could not exceed 
$20\mu$K on a time-scale of a 6-hour rotation period.
This can be met in an orbit close to the second Lagrangian point of the 
Earth-Sun system. The satellite will be always kept out of the Earth shadow and the instruments 
will be shielded from direct illumination by a Sun shield. 

\begin{figure}
  \includegraphics[width=\columnwidth]{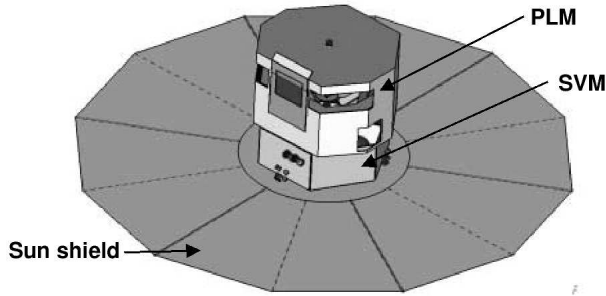}
  \caption{GAIA design with payload module (PLM) and service module (SVM).
  The antenna pointing toward the Earth is on the other side of the Sun shield.
  }
\end{figure}

\begin{figure*}
\includegraphics[width=\textwidth,height=10cm]{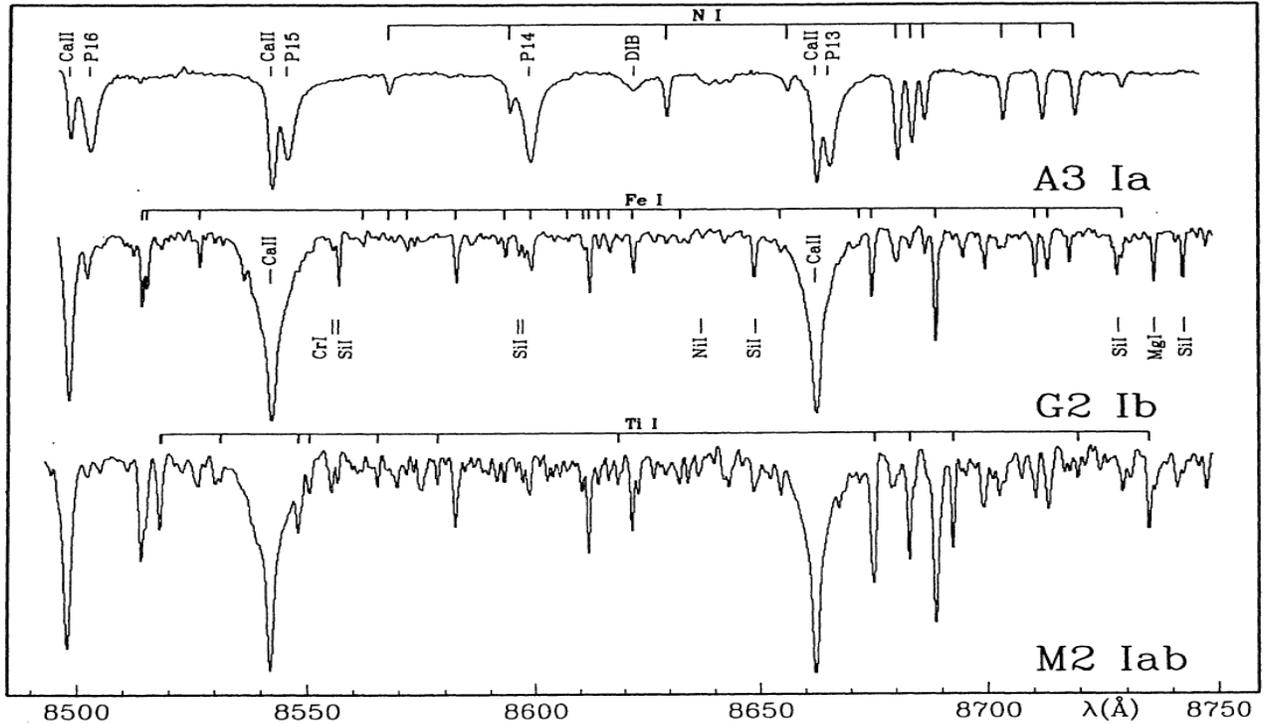}
\caption{Spectra of A, G and M supergiants in the GAIA spectral interval. From Munari (1999).
}
\end{figure*}

\begin{figure}[t]
\includegraphics[angle=-90,width=1.06\columnwidth]{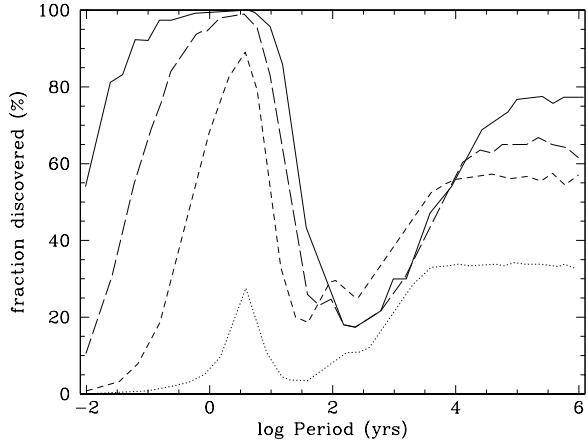}
\caption{Fraction of astrometric and resolved binary stars discovered by GAIA as 
  a function of orbital period and for different magnitude ranges: 
  $10<V<12.5$ (solid), $12.5<V<15$ (long dash), $15<V<17.5$ (short dash), $17.5<V<20$ (dotted
  line). Simulations were done for a distance limited sample ($d < 1 $~kpc). 
  The left maximum marks astrometric systems discovered by a sinusoidal proper 
  motion of the brighter component, the right one is due to systems with resolved components.  
  Adapted from ESA-SCI(2000)4. 
  }
\end{figure}

\begin{figure}[t]
  \includegraphics[angle=-90,width=1.06\columnwidth]{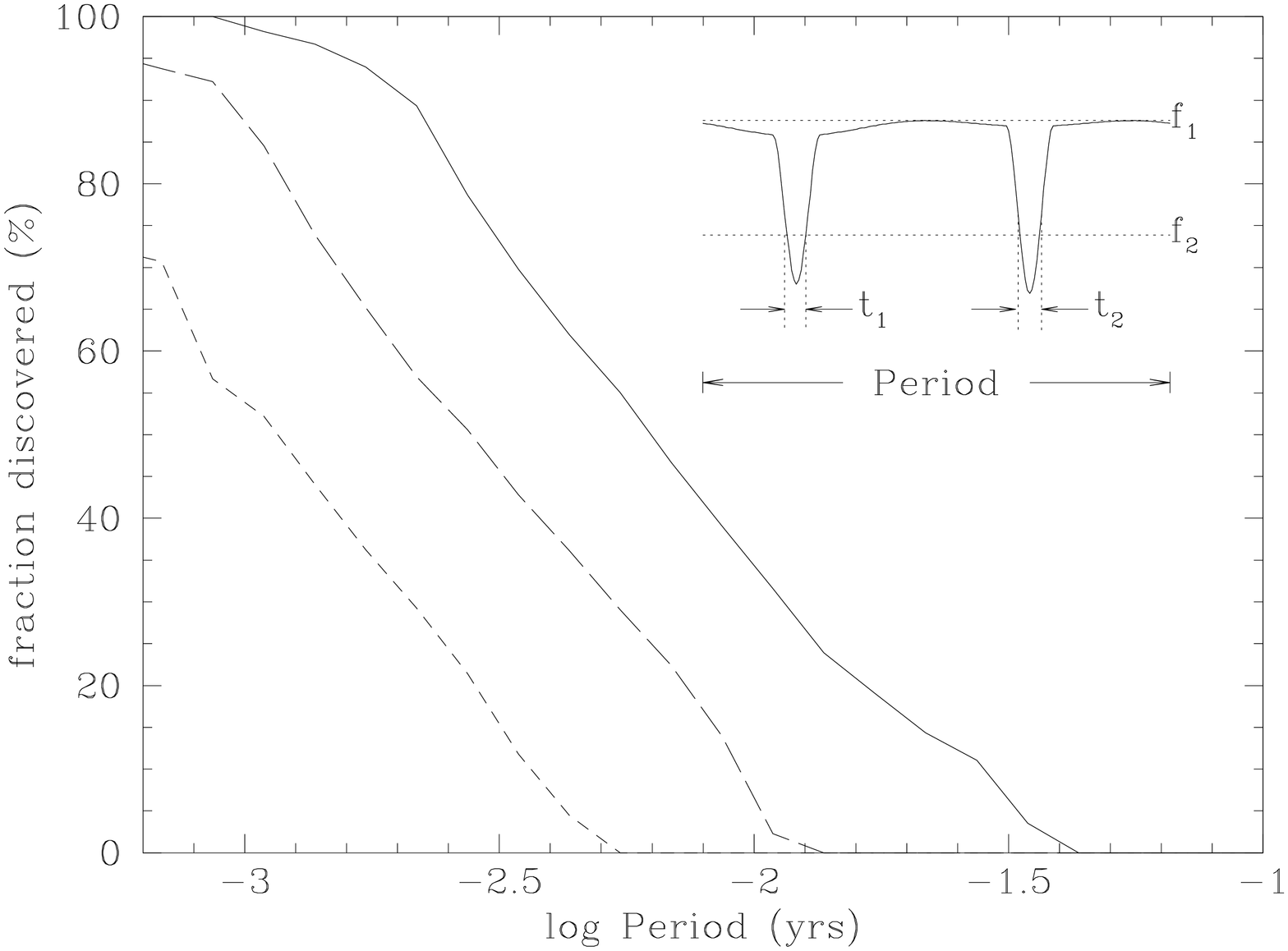}
  \caption{Fraction of binaries discovered by their photometric variability. 
  The components are assumed to be main sequence stars with a total mass 
  of 2 M$_\odot$ and a flat distribution of mass ratios ($0.2 < q < 1.0$).
  A binary gets discovered if its light variation is pronounced enough.
  If the maximum flux level is $f_1$ we assume that the observed flux 
  should be below $f_2$ for at least the time $(t_1+t_2) = 0.08 \times 
  \mathrm{Period}$. The solid curve is for a very accurate photometry 
  where already $f_2/f_1=0.99$ gets detected. The long dashed one is 
  for $f_2/f_1=0.95$ and the short dashed one for the least accurate 
  photometry where $f_2/f_1=0.8$ is needed to bit the noise. 
%
  }
\end{figure}

Astrometric accuracy of $10 \mu$as will be achieved for a V$=15$~mag star after combination of all 
observations during the 5~yr mission. It corresponds to a 10\%\ error in distance at 10~kpc. 
The proper motion error of $10 \mu$as~yr$^{-1}$ equals to a 1~km~s$^{-1}$ error in transverse 
velocity at 20~kpc. So GAIA will be able to directly measure distances and transverse motions with 
a remarkable accuracy and for a representative sample of stars in the Galaxy. 
\medskip

\section{GAIA photometry}

Each star passing the field of view of an astrometric telescope will also be observed in 4 broad 
($\Delta \lambda/ \lambda \sim 0.25$) photometric bands and in white light. Typical errors per 
observation will be between 0.01 and 0.02~mag at V$=18.5$. In addition edges of the focal plane 
of the spectroscopic telescope contain 32 CCDs that will be used for medium band photometry. 
Altogether up to 16 medium band ($\Delta \lambda / \lambda \sim 0.1$) filters will be used. 

Broad band photometry will be excellent to measure general energy distribution of the target. 
Medium band filters are designed to maximize the astrophysical output. For normal stars with 
($V<18$) the goal is to determine $\mathrm{T}_\mathrm{eff}$ to within 50~K and both [M$/$H] 
and $\log g$ to 0.1 dex. The accuracy will be lower for variable stars and for those with 
bright companions.  
\medskip

\section{GAIA spectroscopy}

Any star brighter than $\mathrm{I_c} \sim 16.5$ will have its spectrum observed 
by an on-board spectroscopic telescope. The spectrograph will record spectra 
at a resolving power of $R = 11,500$ in the wavelength range 848--874~nm. 
This range was chosen (Munari 1999) because most GAIA stars will 
be intrinsically red, some of them with a notable interstellar absorption. 
The interval itself is virtually free from telluric absorption, so it will 
be possible to supplement GAIA spectra with pre-- or post--mission ground 
based observations. The wavelength interval contains three strong lines of 
the \ion{Ca}{II} triplet that are present in all dwarfs and giants later 
than B8. In addition, hot stars show Paschen series of Hydrogen, and spectra 
of cool stars contain many metallic lines (Fig.~2). No resonant interstellar 
line is present but intensity of a diffuse interstellar band at 862~nm was found 
to be correlated with reddening (Munari 1999). Peculiar stars are easily detected 
in this wavelength range (Munari 2002).
  
The primary goal of spectroscopy is to measure the radial velocity. The accuracy
has been studied by observations of GAIA-like spectra (Munari et al.\@ 2001a)
and simulations (Katz 2000, Zwitter 2002a). Recapitulation (Munari et al.\@ 2003)
shows that accuracy of 1~km~s$^{-1}$ will be achieved for bright targets 
(see Table~1). Faint stars will suffer both from a low S/N and from crowding 
in the slitless spectrograph focal plane (Zwitter \&\ Henden 2003; Zwitter 2003a). 
So spectra of stars fainter than $\mathrm{I_c} = 16.5$ are not planned to be 
recorded. 

Comparison with observed (Munari \&\ Tomasella 1999) and synthetic (Munari \&\ Castelli
2000; Castelli \&\ Munari 2001) spectra of standard MKK stars in the GAIA spectral interval
could yield other information than radial velocity. Consistency of photometric measurements 
of temperature, gravity and metallicity could be checked. Abundancies of individual 
elements and rotation velocity (Gomboc 2003) could be measured for the brightest stars. 
Moreover any peculiarities, including spectroscopic binarity or multiplicity will be 
readily detected. 

All spectral information will be analyzed on the ground. So successive iterations of 
reduction of each spectrum of a given object could avoid systematic problems, such 
as unknown multiplicity of the target, that will be discovered during the analysis. 
\medskip

\section{Astrometric and resolved binaries}

GAIA will be extremely sensitive to non-linear proper motions. So a representative 
fraction of astrometric binaries with periods from months to tens of years will 
be discovered (Fig.\ 3). GAIA will discover nearly all such binaries brighter than $V=15$
up to a distance of 1~kpc. If the orbital period is shorter than a month the binary 
will be generally too close to discern the sinusoidal component of its proper motion. 
Periods much longer than the mission lifetime will not be recognized because only a 
small fraction of the orbit will be observed. 

Components in binaries within 1~kpc from the Sun and with periods of a hundred thousand years
will be several arc-seconds 
apart and therefore resolved. Since all six space-velocity coordinates of each component 
will be measured, the binary nature of the two stars could be recognized. 

The same reasoning applies to multiple star systems. We conclude that astrometry can 
be used to effectively map multiple systems with orbital periods longer than a month
over the distances up to a few kpc. For many of the stars with orbital periods up to a 
few years a complete orbital solution, including masses of the components, will be 
obtained.
\medskip

\section{Photometric binaries}

Photometric binaries are assumed to be those with an observed photometric 
variability. They include eclipsing binaries, but GAIA photometry will be 
accurate enough to use the reflection effect and discover some non-eclipsing 
systems.
To illustrate the capabilities of GAIA photometry we performed a series 
of simulations of binary light curves. The binary modeling code (Wilson 1998) 
was used via the PHOEBE package (Pr\v{s}a 2003) to simulate a large number 
of photometric light curves. 
The components were assumed to be main sequence stars with a total mass 
of 2~M$_\odot$. The mass ratio was uniformly distributed between 0.2 and 1.0, 
and the orbital plane was observed at a random orientation. Detailed reflection 
effects together with proper limb darkening coefficients were taken into 
account. 

\begin{figure}[t]
  \includegraphics[angle=-90,width=1.06\columnwidth]{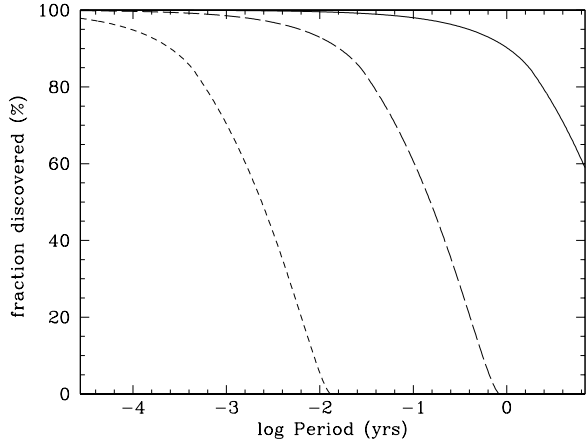}
  \caption{Fraction of single lined spectroscopic binaries discovered by 
  variation in radial velocity of the more luminous component. The components 
  are assumed to be main sequence stars with a total mass of 2 M$_\odot$
  and a flat distribution of mass ratios: $0.2<q<1.0$.
  The curves denote binaries with a $v \sin i$ amplitude of 5 km~s$^{-1}$ (solid),
  20 km~s$^{-1}$ (long dashed) and 80 km~s$^{-1}$ (short dashed line). 
  In the case of accurate radial velocity measurements already 5 km~s$^{-1}$
  variation gets detected and virtually all binaries with periods up to a few 
  years (i.e.\ the mission lifetime) get discovered. If the measurements are more noisy only large 
  velocity amplitudes and so binaries with periods of days or less are 
  recognized.   }
\end{figure}

\begin{figure}[t]
  \includegraphics[angle=-90,width=1.06\columnwidth]{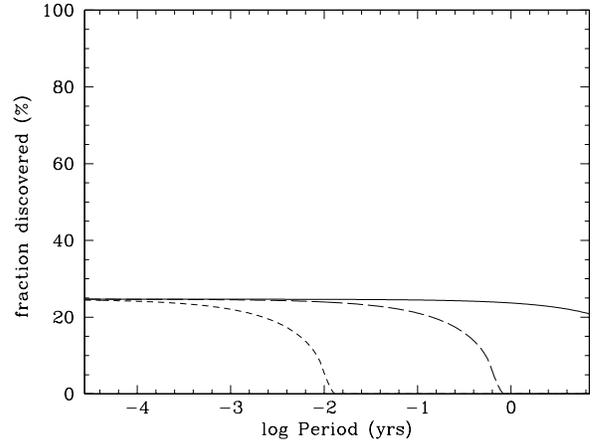}
  \caption{Fraction of discovered double lined spectroscopic binaries. 
  The components are assumed to be main sequence stars  
  with a total mass of 2 M$_\odot$ and a flat distributuon of mass 
  ratios $0.2<q<1.0$. We assume that lines from both stars can be 
  measured if the more massive star has at most twice the luminosity of 
  the fainter one. This limits occurence of double lined systems to 
  $0.8<q<1.0$. The curves denote binaries with a $v \sin i$ amplitude 
  of the more massive component of 5 km~s$^{-1}$ (solid),
  20 km~s$^{-1}$ (long dashed) and 80 km~s$^{-1}$ (short dashed line).
  Interpretation of the required accuracy of radial velocity measurements is 
  similar to Fig.~5. 
  }
\end{figure}

Discovery rate of photometric binaries can be characterized by a fraction of 
the systems that show a certain amplitude of photometric variability. This 
amplitude should not be reached only at one extreme point in the light curve, 
i.e. during a short eclipse, because GAIA will observe the system only 
a limited number of times ($\sim 200$ for middle band and $\sim 82$ for 
broad band filters). Therefore the binary nature of poorly sampled short 
eclipses could not be recognized.
So we assumed that the flux compared to the maximum 
flux should stay below a certain level for at least 8\%\ of the orbital 
period. Therefore $\sim 16$ middle band and $\sim 7$ broad band 
photometric observations fall into these faint phases. This is enough for 
a successful determination of the orbital period. The results (Fig.~4) turn out 
not to depend very much on the length of the considered faint fraction of 
the light curve. The main parameter is the required amplitude of variation. Solid 
curve plots fraction of discovered systems if the flux drops 1\%\ below 
the maximum value for at least 8\%\ of the orbital period. The dashed curves  
are for flux amplitudes of 5 and 20\%. 

It is clear that photometric variability will most easily discover 
binaries that are close to contact. For solar-type components these are 
binaries with orbital periods of a few days or less. The relation of 
course scales with the size of the binary components and only weakly 
depends on the total mass. So systems harbouring giants could be discovered 
at much longer orbital periods. Still, a large luminosity ratio of the 
components, which is typical for this type of systems, will make such discoveries 
difficult.

Comparing Figs.\ 3 and 4 shows that photometric variability becomes 
significant at periods that are just too short for astrometric discovery. 
GAIA photometry will be of high-enough accuracy (see Table~1) to discover 
$\approx 50$~\%\ of all binaries with such short periods. 
GAIA's photometric observations will be obtained in several colours 
simultaneously. So binary light curves could be distinguished 
from intrinsic variability of a single star, e.g.\ due to magnetic spots. 
\medskip

\section{Spectroscopic binaries}

Variation of radial velocity is a very efficient way to discover close 
binaries. In Fig.~5 we plot the situation for a single lined spectroscopic 
binary. The fraction of discovered systems is plotted for different 
amplitudes of the light curve of the more luminous component. 
GAIA will be able to measure single epoch radial velocities with 
$\approx 20$~km~s$^{-1}$ precision for stars brighter than $I_c = 14$. 
So most systems with orbital periods of a few months or less will be 
discovered. In the simulation 
we assumed both stars are main sequence stars with a flat distribution of 
mass ratios in the range $0.2 < q < 1.0$. No intrinsic sources of 
radial velocity variability, like pusations or giant dark spots on rotating 
stellar surfaces, were taken into account. 

Comparison of Figures 3 and 5 shows that the astrometric and spectroscopic 
methods nicely complement each other. A significant fraction of systems 
brighter than $V=15$ could be discovered this way.

Discovery of double lined eclipsing binaries is of utmost importance
(e.g.\ Milone 2003), since only 
for such systems the mass ratio and eventually the masses themselves could 
be accurately determined. In Fig~6 we assumed the same conditions as for 
the single lined binaries, but double lined systems were assumed to have 
the mass ratio in the range $0.8 < q < 1.0$. This means that the fainter 
of the main sequence components still contributes a third of the total light, 
so its lines can be easily measured. The range of preferred orbital 
periods is the same as in the single lined case. Assumption on the intrinsic 
flat distribution of mass ratios in the range $0.2 < q < 1.0$ implies that 
at most a quarter of the systems satisfy the additional condition 
$q>0.8$.
\medskip

\section{Recovery of physical parameters}

\begin{table} \centering
\setlength{\tabnotewidth}{1.0\columnwidth}
\tablecols{6}
\setlength{\tabcolsep}{0.01\columnwidth}
\caption{Distance determination for eclipsing double-lined binary stars  
}
\begin{tabular}{lrrlrl}
\toprule
System      &\multicolumn{5}{c}{MEASURED DISTANCE (pc)}\\ 
Designation &\multicolumn{3}{c}{\ \ Hipparcos parallax}&\multicolumn{2}{c}{Binary analysis\tabnotemark{a}}\\
            &\scriptsize min&aver. &\scriptsize max& & \\
\midrule
V505 Per &\scriptsize 62& 66\ \  &\scriptsize  70 & 59 &$\pm$ 4 \\
V781 Tau &\scriptsize 73& 81\ \  &\scriptsize  91 & 81 & $\pm$ 1\\
UV Leo   &\scriptsize 83 &91\ \  &\scriptsize 103 & 92 & $\pm$ 6\\
V570 Per &\scriptsize 103&117\ \ &\scriptsize 131 &108 &$\pm$ 6 \\
V432 Aur &\scriptsize 100&119\ \ &\scriptsize 146 &124 & $\pm$ 10\\
UW LMi   &\scriptsize 114&129\ \ &\scriptsize 150 &100 & $\pm$ 7\\
GK Dra   &\scriptsize 246&297\ \ &\scriptsize 373 &313 & $\pm$ 14\\
CN Lyn   &\scriptsize 233&362\ \ &\scriptsize 813 &285 & $\pm$ 32\\
OO Peg   &\scriptsize\ \ \  \ \ 304&445\ \ &\scriptsize 840&\ 295 &$\pm$ 17\\
\bottomrule
\tabnotetext{a}{From Munari et al.\@ (2001b), Zwitter et al.\@ (2003), Marrese et al.\@ (2003).}
\end{tabular}

\end{table}

So far we discussed methods to discover binary and multiple stars. Clearly 
one wants to know also what would be the accuracy of physical parameters recovered 
from their analysis. It is sometimes argued that the main role of GAIA is 
to discover interesting systems and that ground-based follow-up observations 
can be used for detailed analysis. While this may be true for a few of the 
most interesting systems it cannot be a general way to proceed. Numbers of 
discovered binary and multiple systems will be huge. Figs.~3-5 imply that 
a significant fraction of binary stars will be discovered at any orbital 
period. The follow up campaigns will be too time consuming so one should 
rely on GAIA data alone. 
A unique feature of GAIA is that any star is observed hundreds of times 
during the mission. The light curves are therefore reasonably well sampled 
(e.g.\ Zwitter 2003b). 

A number of studies (Munari et al.\@ 2001b, Zwitter et al.\@ 2003, 
Marrese et al.\@ 2003) estimated the accuracy of values of physical 
parameters recovered from GAIA-like observations of eclipsing double-lined
spectroscopic binary stars. Hipparcos/Tycho photometry was used as an 
approximation of GAIA photometry, and the Asiago 1.82-m telescope contributed 
Echelle observations in the wavelength range of the GAIA spectrograph. 
Altogether $\sim 20$ systems with solar-type components were observed 
with results on half of them already published. It turns out that masses of 
individual system components can be obtained at an accuracy of 
1-2\%. Similar is also the accuracy of other parameters, except for individual 
stellar radii, which suffer from uneven eclipse sampling. Modeling of 
the total system luminosity yields also the distance. 
The values (Table~2) are consistent with Hipparcos parallaxes, but 
generally more accurate. We note that these results are based on a 
rather noisy Hipparcos photometry, but precision and number of photometric 
bands of GAIA will be much higher.
Astrometric precision of GAIA will be $\sim 100$-times
better than that of Hipparcos. But even so some of the distances to the 
most luminous and remote binary stars will be better determined by system 
modeling than by astrometry. Niarchos \& Manimanis (2003) discussed 
accuracy of recovered physical parameters for near-contact faint systems 
without spectroscopic information. 

GAIA data will be used to solve large numbers of binary and multiple systems 
containing solar and sub-solar stars which generally lack accurate solutions 
in the literature (Andersen 1991). Their absolute placement on the H-R diagram 
and their exact co-evity will allow for extensive testing and improvement 
of the stellar evolution isochrones. It will also close the loop between 
distances from astrometric parallax and eclipsing binary analysis (Wilson 2003). 
Some of the stars in binaries will 
be variable, e.g. of a $\delta$-Scuti type (see Dallaporta et al.\@ 2002, 
Siviero et al.\@ 2003). Direct determination of their masses and radii from 
binary system analysis will improve their physical interpretation. 
\medskip

\section{Project RAVE}

The launch of GAIA is planned for the end of the decade. Radial velocities 
of stars are however a rare commodity (see e.g.\ Munari et el.\@ 2003), 
so earlier results are desired. This is the purpose of the 
{\sl RA}dial {\sl V}elocity {\sl E}xperiment (RAVE), an international 
collaboration led by M.\ Steinmetz which is starting pilot study observations 
already in April 2003. 

The UK Schmidt telescope at AAO equipped with a fiber-optic spectrograph will be used 
to obtain spectra in the same wavelength domain (848--874~nm) as GAIA. In the 
pilot study (2003--2005) about 100,000 stars will be observed, while the main study 
(from 2006) will survey $\sim 35$ million stars in a magnitude limited ($V=16$) 
survey of all sky accessible from AAO, observing $\sim 5$ million stars per year 
(or more if an additional telescope in the northern hemisphere could be used). 

The primary goal of the RAVE experiment is to measure a large sample of 
radial velocities. These can be used to address several galactic kinematic 
issues (Steinmetz 2003).

Pilot study uses the existing hardware of the 6dF survey to secure spectra at 
a resolving power of 8500 during bright unscheduled time. The 6-degree FOV is 
sampled with 150 fibers. Each field of $\sim 130$ stars is exposed for 1~hour. 
For the main study a new fiber optic spectrograph will be built, based 
on the Echidna-style design developed at AAO. It consists of a 2250-spine 
fiber array that covers the full 40~deg$^2$ telescope field of view. Fibers will 
feed a spectrograph operating at a resolving power $\sim 10,000$ in the GAIA 
wavelength range. 

RAVE experiment collects spectra in a way different from GAIA. The latter will 
scan across any star $\sim 100$-times obtaining a low S$/$N spectrum with an effective 
exposure time of 100 seconds on each scan. RAVE's main study will obtain a pointed 30~min 
exposure on each field. A small fraction of stars ($\le 10$~\%) will be 
observed more than once. This means that spectra collected by RAVE will have a good 
S$/$N ratio, but will lack any time variability information. The expected errors on radial 
velocity measurements for RAVE are $\le 2$~km~s$^{-1}$. 

Different observing mode has implications for discovery and analysis of binary and 
multiple stars. Single epoch observations could discover only double-lined binaries. 
Fig.~6 shows that $\sim 10$~\%\ of all binaries observed at random orbital phases 
and with orbital periods up to 10~yrs could be discovered this way. 
Occasional repeated observations of the same target will have the capacity to 
discover a single-lined system. High accuracy of radial velocity measurements implies 
that most of the single-lined spectroscopic binaries with orbital periods 
similar or shorter than the time span of observations could be discovered.  

Discovery of a binary is not enough to assess its physical parameters. GAIA
multi-epoch spectroscopy and photometry will provide good epoch information 
that will generally yield the orbital period and in the lucky cases also a complete 
set of physical parameters. The spectra obtained by RAVE will generally 
not have such a capacity. Velocity separation of the two components is statistically 
linked to the orbital period, and spectra themselves will reveal the spectral type and 
metallicity of either component. But a complete analysis will require dedicated 
follow-up observations. Still the RAVE mission is of extreme importance for binary
studies: contrary to photometric surveys it can discover the vast majority of binaries 
which are not eclipsing. It will provide a great statistical sounding of binary and 
multiple star population 10 years ahead of the more complete and detailed GAIA survey. 
\medskip

\section{Conclusions}

GAIA mission and RAVE experiment are shifting the focus of binary and multiple star 
research. The emphasis moves from a dedicated study of a given system to a standardized 
analysis of a huge sample of such objects. It was estimated (Zwitter 2002b) that GAIA 
may discover $\sim 7 \times 10^6$ eclipsing binaries. At least $\sim 10^4$ of these 
will be double-lined and brighter than $V=15$, permitting a reasonable quality 
determination of their physical parameters. GAIA data will allow for a direct sampling of  
properties of multiple stars at any orbital period, from minutes to millions  
of years. Co-evity of components in such systems will allow unprecedented tests of 
evolutionary theories and formation scenarios. Accuracy of distances obtained from 
analysis of binary systems located at the outskirts of the Galaxy or beyond will rival 
or supersede those obtained by astrometric measurements. The role of the RAVE experiment will 
be to discover a large number of the so far unknown binary and multiple star systems and 
to make a statistical analysis of their properties. 
Still the joy of making a detailed analysis of an individual binary will not be lost. 
GAIA and RAVE will merely point to the really interesting cases that justify such a 
detailed attention.

\acknowledgements
Generous allocation of observing time with the Asiago telescopes has been 
vital to this project. The support of the Slovenian Ministry for Education, 
Science and Sports is acknowledged.


\begin{thebibliography}
\bibitem{} Andersen, J. 1991, A\&A Review, 3, 91
\bibitem{} Bienaym\`{e}, O.,\&\ Turon, C. 2002, eds., {GAIA: a European Space Project},
           (Les Ulis: EDP Sciences), EAS Pub.\ Ser., vol.\ 2
\bibitem{} Castelli, F., \&\ Munari, U. 2001, \aap, 366, 1003
\bibitem{} Dallaporta, S., Tomov T., Zwitter, T., \& Munari, U. 2002, IBVS, 5312
\bibitem{} Gilmore, G., et al.\@ 1998, in SPIE Conf., 3350, ``Astronomical interferometry'', 
   ed. Reasenberg, R. D., 541
\bibitem{} Gomboc, A. 2003,    in ASP Conf.\ Ser., 298, 
   ``GAIA Spectroscopy, Science and Technology'', ed.\ 
   Munari, U. (San Francisco: ASP), 285
\bibitem{} Jordi, C., Babusiaux, C., Katz, D., Portell, J., \& Arenou, F. 2003, UB-SWG-012, 
   http://gaia.am.ub.es/ SWG/Documents/UB\_SWG\_012.pdf
\bibitem{} Katz, D. 2000, PhD Thesis, Univ.\ Paris 7
\bibitem{} Maceroni, C. 2003, this volume
\bibitem{} Marrese, P.M., et al.\ 2003, \aap, submitted
\bibitem{} Milone, E.F. 2003, in ASP Conf.\ Ser., 298, 
   ``GAIA Spectroscopy, Science and Technology'', ed.\ 
   Munari, U. (San Francisco: ASP), 303
\bibitem{} Munari, U. 1999, Baltic Astron., 8, 73
\bibitem{} Munari, U., \&\ Tomasella, L. 1999, \aaps, 137, 521
\bibitem{} Munari, U., \&\ Castelli, F. 2000, \aaps, 141, 141
\bibitem{} Munari, U., Agnolin, P., Tomasella, L. 2001a, Baltic Astron., 10, 613
\bibitem{} Munari, U., et al.\ 2001b. \aap, 378, 477
\bibitem{} Munari, U. 2002, in ASP Conf.\ Ser., 279, 
   ``Exotic Stars as Challenges to Evolution'', eds.\ 
   Tout, C. A. \&\ Van Hamme, W. 
   (San Francisco: ASP), 25 
\bibitem{} Munari, U. 2003, ed., ASP Conf.\ Ser., 298, 
   ``GAIA Spectroscopy, Science and Technology''
\bibitem{} Munari, U., Zwitter, T., Katz, D., \&\ Cropper M. 2003, 
   in ASP Conf.\ Ser., 298, 
   ``GAIA Spectroscopy, Science and Technology'', ed.\ 
   Munari, U. (San Francisco: ASP), 275
\bibitem{} Niarchos, P.G., \& Manimanis, V.N. 2003, \aap, accepted
\bibitem{} Perryman, M. A. C., et al.\@ 2001, \aap, 369, 339
\bibitem{} Pr\v{s}a, A. 2003, in ASP Conf.\ Ser., 298, 
   ``GAIA Spectroscopy, Science and Technology'', ed.\ 
   Munari, U. (San Francisco: ASP), 457
\bibitem{} Siviero, A., et al.\@ 2003, \aap, submitted
\bibitem{} Steinmetz, M. 2003, in ASP Conf.\ Ser., 298, 
   ``GAIA Spectroscopy, Science and Technology'', ed.\ 
   Munari, U. (San Francisco: ASP), 381
\bibitem{} Strai\v{z}ys, V. 1999, ed., ``GAIA'', Baltic Astron., 8, 1
\bibitem{} Vansevi\v{c}ius, V., Ku\v{c}inskas, A., \& Sud\v{z}ius, J. 
   2002, eds., Kluwer, ``Census of the Galaxy: Challenges for Photometry
   and Spectrometry with GAIA''
\bibitem{} Wilson, R. E. 1998, Computing Binary Star Observables,
             Univ. of Florida Astronomy Dept.
\bibitem{} Wilson, R. E. 2003, in ASP Conf.\ Ser., 298, 
   ``GAIA Spectroscopy, Science and Technology'', ed.\ 
   Munari, U. (San Francisco: ASP), 313
\bibitem{} Zwitter, T. 2002a, \aap, 386, 748
\bibitem{} Zwitter, T. 2002b, in ASP Conf.\ Ser., 279, 
   ``Exotic Stars as Challenges to Evolution'', eds.\ 
   Tout, C. A. \&\ Van Hamme, W. 
   (San Francisco: ASP), 31
\bibitem{} Zwitter, T., \&\ Henden A. 2003, in ASP Conf.\ Ser., 298, 
   ``GAIA Spectroscopy, Science and Technology'', ed.\ 
   Munari, U. (San Francisco: ASP), 489 
\bibitem{} Zwitter, T. 2003a, in ASP Conf.\ Ser., 298, 
   ``GAIA Spectroscopy, Science and Technology'', ed.\ 
   Munari, U. (San Francisco: ASP), 493
\bibitem{} Zwitter, T. 2003b, in ASP Conf.\ Ser., 298, 
   ``GAIA Spectroscopy, Science and Technology'', ed.\ 
   Munari, U. (San Francisco: ASP), 329
\bibitem{} Zwitter, T., et al.\ 2003, \aap, accepted, in print (astro-ph/0303573)
\end{thebibliography}
\end{document}